\documentclass[amssymb,prd,superscriptaddress,aps,nofootinbib,twocolumn,showpacs]{revtex4-1}
\usepackage{graphicx, epsfig, amssymb} 
\usepackage{amsmath, amsfonts}
\usepackage{bm} 
\usepackage[breaklinks]{hyperref}
\usepackage{color}
\usepackage{enumerate}
\usepackage{units}

\def\be{\begin{equation}}
\def\ee{\end{equation}}
\def\beq{\begin{eqnarray}}
\def\eeq{\end{eqnarray}}

\begin{document}

\title{Spin evolution of a proto-neutron star}


\author{Giovanni Camelio}\email{giovanni.camelio@roma1.infn.it} \affiliation{Dipartimento di Fisica, ``Sapienza''
  Universit\`a di Roma \& Sezione INFN Roma1, P.A. Moro 5, 00185, Roma, Italy.}

\author{Leonardo Gualtieri}\email{leonardo.gualtieri@roma1.infn.it} \affiliation{Dipartimento di Fisica, ``Sapienza''
  Universit\`a di Roma \& Sezione INFN Roma1, P.A. Moro 5, 00185, Roma, Italy.}

\author{Jos\'e A. Pons}\email{jose.pons@ua.es} \affiliation{Departament de F\'isica Aplicada, Universitat d’Alacant,
  Ap. Correus 99, 03080 Alacant, Spain.}

\author{Valeria Ferrari}\email{valeria.ferrari@roma1.infn.it} \affiliation{Dipartimento di Fisica, ``Sapienza''
  Universit\`a di Roma \& Sezione INFN Roma1, P.A. Moro 5, 00185, Roma, Italy.}

\begin{abstract}
We study the evolution of the rotation rate of a proto-neutron star, born in a core-collapse supernova, in the first
seconds of its life. During this phase, the star evolution can be described as a sequence of stationary configurations,
which we determine by solving the neutrino transport and the stellar structure equations in general relativity. We
include in our model the angular momentum loss due to neutrino emission.  We find that the requirement of a rotation
rate not exceeding the mass-shedding limit at the beginning of the evolution implies a strict bound on the rotation
rate at later times. Moreover, assuming that the proto-neutron star is born with a finite ellipticity, we determine the
emitted gravitational wave signal, and estimate its detectability by present and future ground-based interferometric
detectors.
\end{abstract}


\maketitle

\section{Introduction}\label{sec:intro}
When a supernova explodes, it leaves a hot, lepton-rich and (presumably) rapidly
rotating remnant: a proto-neutron star (PNS).  In the early stages of its
evolution, the PNS cools down and loses its high lepton content, while its radius and  rotation
rate decrease. In this phase, a huge amount of energy and of angular momentum is
released, mainly due to neutrino emission~\cite{Burrows:1986me,Keil:1995hw,Pons:1998mm}.
A fraction of this energy is expected to be emitted in the gravitational wave
channel; indeed, as a consequence of the violent collapse non-radial oscillations can be
excited, making PNSs promising sources for present and future gravitational
detectors~\cite{Ferrari:2002ut,Ott:2008wt,Burgio:2011qe,Fuller:2015lpa}.

In the first tenths of seconds after its birth, the PNS is turbulent and
characterized by large instabilities. During the next tens of seconds, it
undergoes a more quiet, ``quasi-stationary'' evolution (the Kelvin-Helmholtz phase),
which can be described as a sequence of equilibrium
configurations~\cite{Burrows:1986me,Pons:1998mm}.  In this article, we study the
evolution of the rotation rate of a PNS during this quasi-stationary,
Kelvin-Helmholtz phase.  An accurate modeling  of this phase is needed, for
instance, to compute the frequencies of the PNS gravitational wave emission.
Moreover, it provides a link bewteen supernova explosions, a phenomenon
which is still not fully understood, and the properties of the
observed population of young pulsars. Current models of the
evolution of progenitor stars~\cite{Heger:2004qp}, combined with numerical
simulations of core collapse and explosion (see
e.g.~\cite{Thompson:2004if,Ott:2005wh,Hanke:2013jat,Couch:2014kza,Nakamura:2014fxa}), do not
allow sufficiently accurate estimates of the expected rotation rate of newly
born PNSs; they only show that the minimum rotation period
at the onset of the Kelvin-Helmoltz phase can be as small as few ms, if the spin rate of
the progenitor is sufficiently high.
%
On the other hand, astrophysical observations of young pulsar populations
(see~\cite{Miller:2014aaa} and references therein) show typical periods
$\gtrsim~100$ ms.

The quasi-stationary evolution of a PNS driven by neutrino transport in a spherically symmetric space-time
has been extensively studied in the past, quite often adopting  an equation of state (EoS) obtained within a finite-temperature,
field-theoretical model solved at the mean field level~\cite{Pons:1998mm,Pons:2000xf,Pons:2001ar,Roberts:2012zza}. 
This approach yields a sequence of thermodynamical profiles
describing the structure and the early evolution of a non-rotating PNS.
A different approach has been used in~\cite{Burgio:2011qe}, where an EoS obtained within a
finite-temperature many-body theory approach was employed, but the neutrino
transport equations were not explicitly solved (a set of entropy and lepton
fraction profiles were adopted, having the same qualitative behaviour as those
of~\cite{Pons:1998mm}). We also mention that the non-radial oscillations of the
quasi-stationary configurations obtained with these different approaches have
been studied in~\cite{Ferrari:2002ut,Ferrari:2003qu,Burgio:2011qe}, where
the quasi-normal mode frequencies of the gravitational waves emitted
in the early PNS life were computed.

The evolution of rotating PNSs has been studied in~\cite{Villain:2003ey}, where
the  thermodynamical profiles obtained in~\cite{Pons:1998mm}
for a non-rotating PNS were employed as effective one-parameter EoSs;
the rotating configurations were
obtained using the non-linear BGSM code~\cite{Gourgoulhon:1999vx} to solve
Einstein's equations. A similar approach has been followed
in~\cite{Martinon:2014uua}, which used  the profiles of~\cite{Pons:1998mm}
and~\cite{Burgio:2011qe}. The main limitations of these works are the
following.
\begin{itemize}
\item The evolution of the PNS rotation rate is due not only to the change in the moment of inertia (i.e., to the
  contraction), but also to the angular momentum change due to neutrino emission~\cite{Epstein:1978md}.  This was
  neglected in~\cite{Villain:2003ey}, and described with an heuristic formula in~\cite{Martinon:2014uua}.
\item As we shall discuss in this paper, when the PNS profiles describing a non-rotating
star are treated as an effective EoS, one can obtain
configurations which are unstable to radial perturbations.
\end{itemize}
In this article, we study the quasi-stationary evolution of a spherically symmetric PNS, solving the relativistic
equations of neutrino transport and of stellar structure. The details of our code will be discussed
in~\cite{Camelio:2016}, where it will be applied to more recent EoSs. Here, we employ the same EoS used
in~\cite{Pons:1998mm} (i.e.  GM3~\cite{Glendenning:1991es}), to study the spin evolution of the PNS in its first tens of
seconds of life.  To model an evolving, rotating PNS, we use the profiles of entropy per baryon and lepton fraction
$s(a)$, $Y_L(a)$ ($a$ is the number of baryons enclosed in a sphere passing through the point considered) obtained with
our evolution code which describes a non-rotating PNS. Our approach is different from that used in
\cite{Villain:2003ey}, as will be discussed in detail in Sec~\ref{sec:thermorot}. In order to determine the PNS spin
evolution, we model the evolution of angular momentum (due to neutrino emission) using Epstein's
formula~\cite{Epstein:1978md}. We also discuss the gravitational wave emission which could be associated with this
process.

The plan of the paper is as follows. In Sec.~\ref{sec:evolution} we briefly
describe our approach to model the PNS evolution in its quasi-stationary phase.
In Sec.~\ref{sec:rotation} we describe our model of a rotating PNS. In
Sec.~\ref{sec:results} we show our results, and in Sec.~\ref{sec:conclusions} we
draw our conclusions. The details of the slowly rotating model are described in
Appendix~\ref{app:HT}.

\section{Early evolution of a proto-neutron star}\label{sec:evolution}
The quasi-stationary, Kelvin-Helmholtz phase of a PNS starts few hundreds of ms
after the core bounce\cite{Burrows:1986me,Pons:1998mm}. This phase consists of
two evolutionary stages. In the first few tens of seconds, neutrinos diffuse from
the low-entropy core to the high-entropy envelope, deleptonizing the core and
increasing its temperature. In the second phase, the star is lepton poor but
hot, the entropy gradient is smoothed out, and thermally produced neutrinos
cool down the PNS. After about one minute, the star becomes transparent to neutrinos
and can be considered as a ``mature'' neutron star, with a radius of
$\sim10-15$ km and a temperature $<1$ MeV.

The PNS evolution in the Kelvin-Helmholtz phase can be considered as a sequence
of quasi-stationary configurations, because the hydrodynamical timescale
is much smaller than the evolution timescale.
Following~\cite{Pons:1998mm}, we model this phase by solving the general
relativistic neutrino transport equations coupled with the structure equations,
assuming spherical symmetry. In each quasi-stationary configuration, the
spacetime metric has the form
\begin{equation}
ds^2=-e^{\phi(r)}dt^2+e^{\lambda(r)}dr^2+r^2(d\theta^2+\sin^2\theta
d\varphi^2)\,, \label{metricinsidespher}
\end{equation}
where $\phi(r)$ and $\lambda(r)$ are radial functions, obtained by solving the
Tolman-Oppenheimer-Volkov (TOV) equations (in this paper we use geometrized
units, in which $c=G=1$).  The perfect fluid of the star is described by the
stress-energy tensor $T_{\mu\nu}=(\epsilon+p)u_\mu u_\nu+pg_{\mu\nu}$, where
$u^\mu=(e^{-\phi/2},0,0,0)$ is the fluid four-velocity and $\epsilon,p$ are the
energy density and pressure of the fluid, respectively. The gravitational mass
inside a radius $r$ is $m(r)=r(1-e^{-\lambda})/2$. At the surface of the star,
$r=R$, the pressure vanishes and the metric matches with the exterior
Schwarzschild metric, with $M=m(R)$ as the gravitational mass of the star. We also
define the baryon number inside a radius $r$,
\begin{equation}
a(r)=4\pi\int_0^re^{\lambda/2}n_br'^2dr'\,,
\end{equation}
where $n_b$ is the baryon number density. The position inside the star can be described either by the coordinate $r$, or
by the enclosed baryon number $a$. We also define the rest-mass density $\rho=m_n n_b$ ($m_n$ is the neutron mass), and
the baryon mass of the star $M_b=m_na(R)$. We will use $M_b=1.60~M_\odot$, which corresponds, in the calculations of
this paper, to a gravitational mass of $1.55~M_\odot$ at $200$ ms from the core bounce, which reduces to
$\sim1.4~M_\odot$ in the first ten seconds of life of the PNS. 

Since the PNS has a temperature of several MeV, its EoS is non-barotropic, and
can be written as
\begin{equation}
\epsilon=\epsilon(p,s,\{Y_i\}_i)\,,
\end{equation}
where $s$ is the entropy per baryon, and $Y_i=n_i/n_b$ is the fraction of the $i$-th specie, with number density
$n_i$. Assuming that the matter is in beta equilibrium, the dependence on the composition $\{Y_i\}_i$ can be cast into a
dependence on only the electron-type lepton fraction $Y_L=n_L/n_b$. Different choices of thermodynamical variables are
possible, for instance replacing the entropy per baryon $s$ with the temperature $T$.  In this paper, we use the
finite-temperature EoS GM3 of Glendenning and Moszkowski~\cite{Glendenning:1991es}, obtained within a field-theoretical
model solved at the mean field level, where the interactions between baryons are mediated by the exchange of mesons; it
contains only nucleonic degrees of freedom.  This is the same EoS employed in~\cite{Pons:1998mm}; we consider the case of matter composed by electrons, protons and neutrons. More recent EoSs,
based on a many-body theory approach, will be considered in a future work~\cite{Camelio:2016}.

In order to solve the TOV equations, we need to know, at each point, the energy
density as a function of the pressure; thus, we need to know the EoS and the
thermodynamical profiles $s(a)$, $Y_L(a)$, which are obtained by solving the
transport equations
\begin{align}
\frac{\partial Y_L}{\partial t}+\frac{\partial(4\pi
e^{\phi/2}r^2F_\nu)}{\partial a}&=0\,,\label{transpeq1}\\
T\frac{\partial s}{\partial t}+\mu_\nu\frac{\partial Y_L}{\partial t}+e^{-\phi/2}
\frac{\partial(4\pi e^{\phi}r^2H_\nu)}{\partial a}&=0\,,\label{transpeq}
\end{align}
where $F_\nu$ and $H_\nu$ are, respectively, the neutrino number and energy fluxes
\begin{align}
\label{eq:Fnu}
F_\nu={}&-\frac{\mathrm e^{-\frac{\lambda+\phi}2}T^2}{6\pi^2\hbar^3}
\Bigg[D_3\frac{\partial(T\mathrm e^{\phi/2})}{\partial r}+(T\mathrm
e^{\phi/2})D_2\frac{\partial\eta}{\partial r}\Bigg]\,,\\
\label{eq:Hnu}
H_\nu={}&-\frac{\mathrm e^{-\frac{\lambda+\phi}2}T^3}{6\pi^2\hbar^3}
\Bigg[D_4\frac{\partial(T\mathrm e^{\phi/2})}{\partial r}+(T\mathrm
e^{\phi/2})D_3\frac{\partial\eta}{\partial r}\Bigg]\,,
\end{align}
where $\eta=\mu_\nu/T$ is the electron-type neutrino degeneracy parameter and
$D_n$ are the neutrino diffusion coefficients, which are computed assuming the diffusion
approximation~\cite{Pons:1998mm}.
In order to preserve causality and stabilize the code in the semi-transparent regions near the
PNS surface, we apply a flux-limiter~\cite{levermore1981flux}.

\begin{figure}[ht]
\begin{center}
\leftskip-3em
\epsfig{file=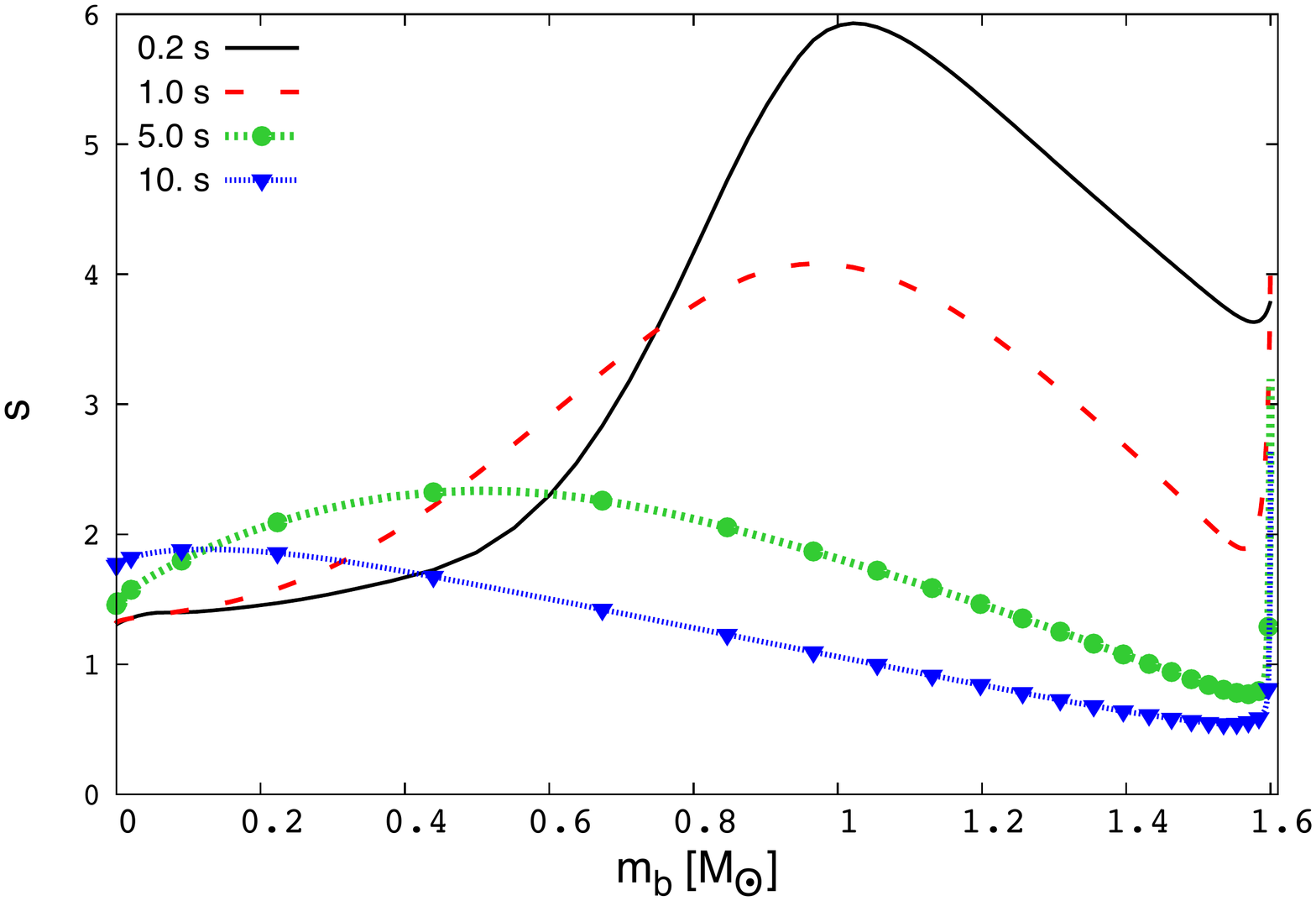,width=7.5cm,angle=270,clip=true}
\caption{Entropy per baryon as a function of the enclosed baryon mass
$m_b=m_na$
at $t=0.2,1,5,10$ s ($k_\mathrm B=1$).
\label{fig:st}}
\end{center}
\end{figure}

\begin{figure}[ht]
\begin{center}
\leftskip-3em
\epsfig{file=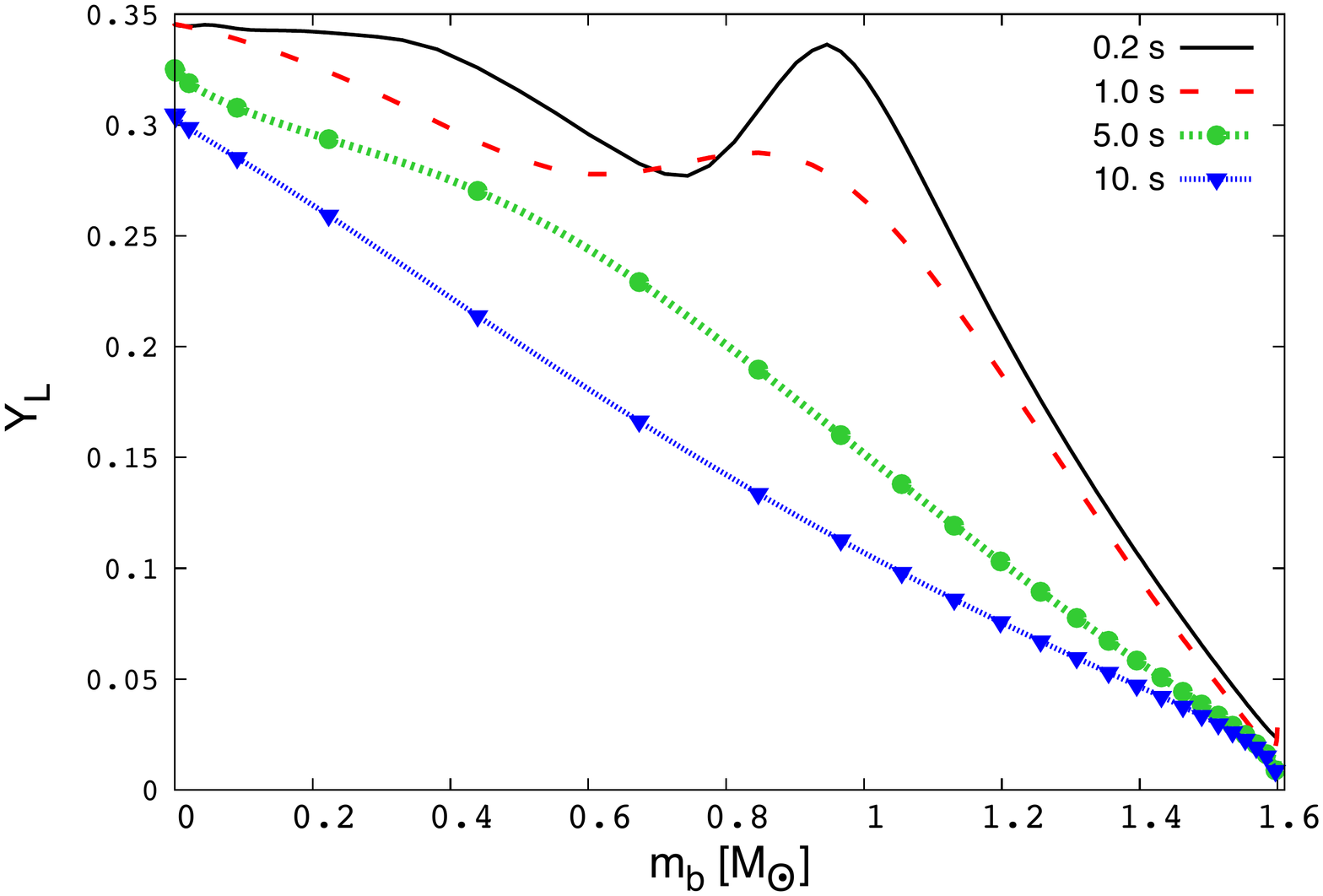,width=7.5cm,angle=270,clip=true}
\caption{Electron-type lepton fraction as a function of the
enclosed baryon mass  at $t=0.2,1,5,10$ s.
\label{fig:YLt}}
\end{center}
\end{figure}

Our code evolves the PNS by iteratively solving, at each time-step, (i) the transport
equations~\eqref{transpeq1} and \eqref{transpeq} using an implicit scheme and
(ii) the TOV equations by relaxation method. The time evolution keeps the baryon mass
$M_b$ constant, and provides, at each time-step, a quasi-stationary
configuration of the (non-rotating) star, described by the profiles of all the
thermodynamical quantities ($p$, $\epsilon$, $n_b$, $s$, $Y_L$, etc.) as
functions of $a$ (or of $r$).
We start our integration at $200$ ms from core bounce. The initial profiles (which are the same
employed in~\cite{Pons:1998mm}) are the result of core-collapse simulations~\cite{wilson1989nuclear}.
In Figures~\ref{fig:st} and \ref{fig:YLt} we
show the evolutionary profiles of the entropy per baryon $s$ and the
electron-type lepton fraction $Y_L$ as functions of the enclosed baryon mass.  We have checked that the total energy and
lepton number are conserved during the evolution within a few percent in the early stages of the evolution, and with more
accuracy in later stages. We remark that this
error can be significantly reduced by reducing the timestep; however, this accuracy is sufficient for the aims of this
work. Our code will be described in detail in a future work \cite{Camelio:2016}. 

%
Recently, different approaches have been applied in the study of the PNS
evolution (see e.g.~\cite{Roberts:2012zza}), in which the neutrino spectrum is described with greater accuracy by means
of multi-group codes. However, since in this work we are not interested in the details of the neutrino emission, we
prefer to employ a simpler and faster energy-averaged approach (as
in~\cite{Pons:1998mm}). As mentioned above, our code also employs a 
flux-limiter~\cite{levermore1981flux}, which makes it difficult to establish the precise location of the
neutrinosphere. Both the neutrinosphere and the neutrino spectra are better determined with more complex core-collapse
codes, which however are far slower, while our PNS code is suitable to run for longer evolution times.  Typical
core-collapse codes run for at most $500$~ms after core bounce, whereas we can easily explore the first minute of PNS
life, at the end of which the star becomes neutrino-transparent.


\section{A model of rotating proto-neutron stars}\label{sec:rotation}

\subsection{Slowly rotating stars in general relativity}\label{sec:HT}
We model a rotating PNS using the perturbative approach of Hartle and
Thorne~\cite{Hartle:1967he,Hartle:1968si} (see also~\cite{Benhar:2005gi}).  The
rotating star is described as a stationary perturbation of a spherically
symmetric background, for small values of the angular velocity $\Omega=2\pi\nu$,
i.e., $\nu\ll\nu_{ms}$ ($\nu_{ms}$ is the mass-shedding
frequency, at which the star starts losing mass at the equator, see
Sec.~\ref{sec:ms}).  As shown in~\cite{Martinon:2014uua}, this ``slow rotation''
approximation is reasonably accurate for rotation rates up to $\sim0.8$ of the
mass-shedding limit, providing values of mass, equatorial radius and moment of
inertia which differ by $\lesssim0.5\%$ from those obtained with fully
relativistic, nonlinear simulations.
In our approach we assume uniform rotation; PNSs are expected to
have a significant amount of differential rotation at
birth~\cite{janka1989hydrostatic} which, however, is likely to be removed by viscous mechanisms, such as, for
instance, magnetorotational instability~\cite{Mosta:2015ucs}, in a fraction of a second.

This work should be considered as a first
step towards a more detailed description of rotating PNSs, in which we shall
include differential rotation.

The spacetime metric, up to third order in $\Omega$, can be written as
\begin{align}
\label{metricHartle}
\nonumber ds^2 &= -e^{\phi(r)} [ 1 + 2h_0(r) + 2h_2(r)P_2(\mu) ] dt^2 \\
\nonumber      &+ e^{\lambda (r)}\left[ 1 + \frac{ 2m_0(r) +
2m_2(r)P_2(\mu) }{ r - 2M(r) } \right]dr^2 \\
               &+ r^2 [ 1 + 2k_2(r)P_2(\mu) ] \\
               \label{ds2}
\nonumber      &\times[ d\theta^2 + \sin^2\theta \{ d\phi - [\omega (r)
+ w_1(r) + w_3(r)P^{'}_3(\mu)]dt \}^2 ]
\end{align}
where $\mu=\cos\theta$ and $P_n(\mu)$ is the Legendre polynomial of order $n$,
the prime denoting the derivative with respect to $\mu$.  The perturbations of
the non-rotating star are described by the functions $\omega$ (of $O(\Omega)$),
$h_0$, $m_0$ and $h_2$, $m_2$, $k_2$ (of $O(\Omega^2)$), and $w_1$, $w_3$ (of
$O(\Omega^3)$).  The energy-momentum tensor is
\begin{equation}
   T^{\mu\nu} = ({\cal E} + {\cal P})u^{\mu}u^{\nu} + {\cal P}g^{\mu\nu}
\end{equation}
where $g_{\mu\nu}$, $u^{\mu}$ are the metric and four-velocity in the rotating
configuration, and we denote by calligraphic letters thermodynamical quantities
(energy, density and pressure) in the rotating star. An element of
fluid, at position $(r,\theta)$ in the non-rotating star, is displaced by
rotation to the position
\begin{equation}
\label{displ1}
\bar{r}= r + \xi(r,\theta)\,,
\end{equation}
where $\xi(r,\theta)=\xi_0(r) + \xi_2(r)P_2(\mu) +O(\Omega^4)$ is the Lagrangian displacement.

In the Hartle-Thorne approach, one assumes that if the fluid element of the non-rotating star has
pressure $P$ and energy density $\epsilon$, the displaced fluid element of the rotating star has the same values of
pressure and energy density.  In other words, the Lagrangian perturbations of the thermodynamical quantities $\epsilon$,
$P$ vanish (see~\cite{Hartle:1967he}, eq.\,$(6)$); the modification of these quantities is only due to the
displacement (\ref{displ1}):
\begin{equation}
  \delta \epsilon(r,\theta) = - \frac{d\epsilon}{dr}\xi(r,\theta)~~~\,,~~~~
\delta P(r,\theta) = - \frac{dP}{dr}\xi(r,\theta)\,.  \label{epspert}
\end{equation}
We remark that as long as we neglect terms of $O(\Omega^4)$,
$\delta \epsilon(r,\theta)\simeq\delta \epsilon(\bar r,\theta)$.

Einstein's equations, expanded in powers of $\Omega$ and in Legendre polynomials, can be written as a set of ordinary
differential equations for the perturbation functions; these equations are summarized in Appendix~\ref{app:HT}. For each
value of the central pressure $p_c$ (or, equivalently, of the central energy density $\epsilon_c$) and of the rotation
rate $\Omega$, the numerical integration of the perturbation equations yields the perturbed functions, and then the
values of the multipole moments of the star (in particular, the mass $M$ and the angular momentum $J$), and of its
baryonic mass $M_b$. These quantities can be written as $M=M^{(0)}+\delta M$, $J=\delta J$, $M_b=M_b^{(0)}+\delta M_b$,
etc., where the quantities with superscipt $^{(0)}$ refer to the non-rotating star with central pressure $p_c$, and the
quantities with $\delta$ are the corrections due to rotation.

Given a non-rotating star with central pressure $p_c$ and baryon mass $M_b$, the
rotating star (with spin $\Omega$) with the same central pressure has a baryon mass
$M_b^{(0)}+\delta M_b$, which is generally larger than $M_b$. Therefore, a
rotating star with same baryon mass $M_b$ as the non-rotating one, has
necessarily a smaller value of the central pressure, $p_c+\delta p_c$, with $\delta p_c <0$
(this is not surprising: when a star is set into rotation, its central pressure
decreases).

We mention that in ~\cite{O_Connor+Ott.2010} 
the neutrino transport equations for a rotating star in general relativity have
been solved by using an alternative approach. 
In this approach (which is believed to be
accurate for slowly rotating stars~\cite{O_Connor+Ott.2010}) the structure and
transport equations for a spherically symmetric star are modified by adding a
centrifugal force term, to include the effect of rotation.

\subsection{Including the thermodynamical profiles}\label{sec:thermorot}
In order to integrate the structure equations of a cold neutron star we need
to assign an equation of state which,
in the case of PNSs,  is non-barotropic, i.e.
$\epsilon=\epsilon(p,s,Y_L)$, thus we
also need to know the profiles of entropy and lepton fraction throughout the
star. As discussed in Section~\ref{sec:evolution}, these
profiles are obtained by our evolutionary code for spherical, non-rotating PNS
at selected values of time.

The non-rotating profiles can be used to compute the structure of a  rotating PNS
in different ways. A possible approach is the following.

Let us consider  a
spherical PNS with baryon mass $M_b$  at a given value of the
evolution time $t$. The numerical code discussed in Sec.~\ref{sec:evolution}
provides the functions $p(a)$, $\epsilon(a)$, $s(a)$, $Y_L(a)$, where we remind that
$a$ is the enclosed baryon number.
If we replace the inverse function of
$p(a)$ into the non-barotropic EoS, we obtain an ``effective barotropic EoS'',
$\tilde\epsilon(p)=\epsilon(p,s(a(p)),Y_L(a(p))$, which can be used to solve the TOV
equations for the spherical configuration to  which we  add the
perturbations due to rotation, according to Hartle's procedure.
Since the rotating star must have the same baryon mass as the spherical star, 
one can proceed  as follows: 
(i) solve the TOV equations for a spherically symmetric star with central pressure $p_c+\delta p_c$;
(ii) solve the perturbation equations for a chosen
value of the rotation rate, to determine the actual baryon mass of the rotating star with same central pressure;
(iii) iterate these two steps modifying $\delta p_c$  until the baryon mass coincides with the assigned  value $M_b$.
This approach was used in~\cite{Villain:2003ey}, where the
rotating star was modeled solving the fully non-linear Einstein equations.

However, this procedure has some relevant drawbacks.  Indeed, during the first second after bounce the star is very
weakly bound, and it may happen that the procedure above yields $\delta p_c>0$, which indicates that these
configurations are in the unstable branch of the mass-radius diagram.  We think that this is caused by the unphysical
treatment of the thermodynamical profile (effectively, as a barotropic EoS).

This problem did not occur in the simulations of~\cite{Villain:2003ey} because the authors considered a different,
stable branch of the mass-radius curve corresponding to the ``effective'' EoS $\tilde\epsilon(p)$, at much lower
densities. Indeed, for $t\lesssim0.5$ s, at the center of the star they had $n_b\sim\unit[10^{-2}]{fm^{-3}}$ (i.e.,
rest-mass density $\rho\sim\unit[10^{13}]{g/cm^{3}}$), which corresponds to the outer region of the star modeled
in~\cite{Pons:1998mm}.  When the central density is so low, only a small region of the star is described by the GM3 EoS;
the rest is described by the low-density EoS used to model the PNS envelope, which does not yield unstable
configurations.

\begin{figure*}[ht]
\begin{center}
\epsfig{file=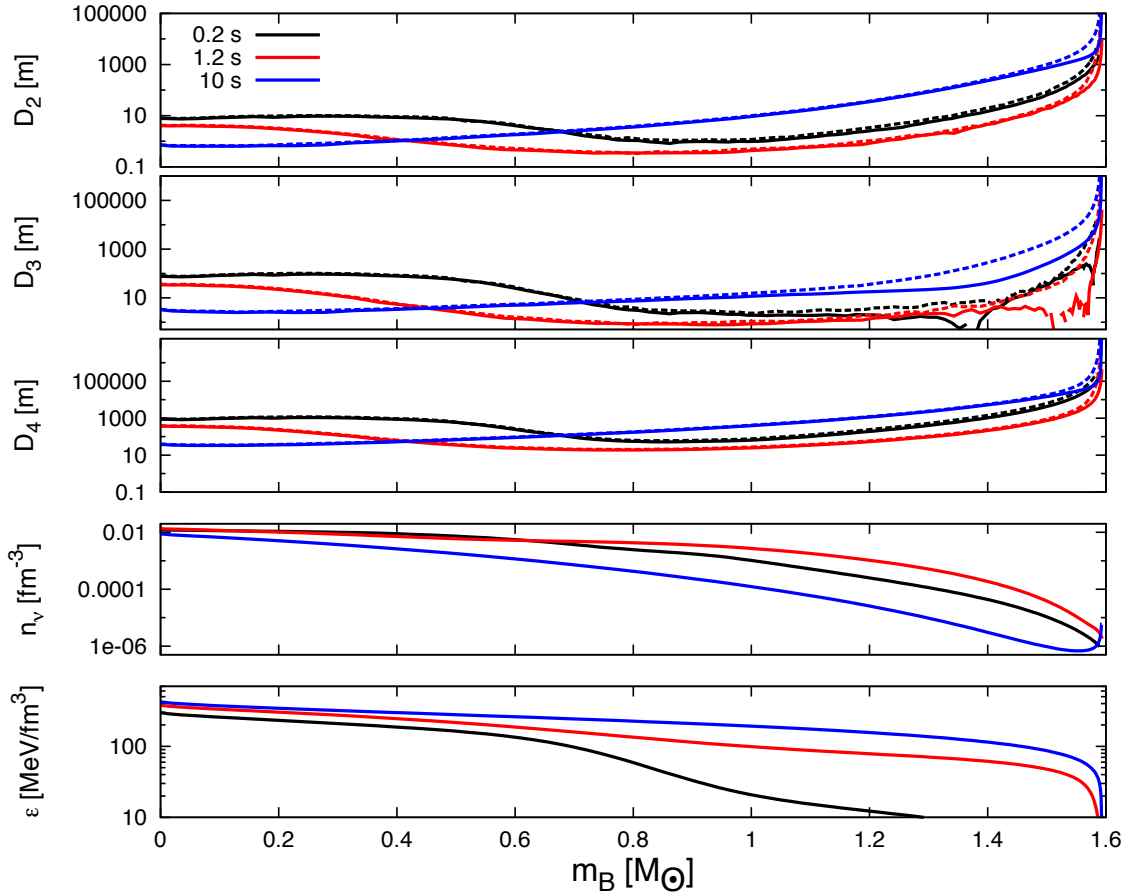,width=17.cm,clip=true}
\caption{Neutrino diffusion coefficients $D_n$ ($n=2,3,4$), as functions of the enclosed baryon mass, computed using the
  density and thermodynamical profiles of the non-rotating (solid line) and rotating (dashed line) configurations, at
  $t=0.2$ s, $t=1.2$ s and $t=10$ s (upper and middle panels).  Profiles of neutrino number density and energy density
  (lower panels).  We assume $M_b=1.6\,M_{\odot}$ and that the angular momentum is the
  maximum allowed $J_{in}=J_{max}$ (for $J_{in}>J_{max}$, the PNS reaches the mass-shedding limit during its evolution,
  see Sec.~\ref{sec:resspin}).\label{fig:compare}}
\end{center} \end{figure*}

Since we want to model the PNS consistently with the
evolutionary models given in \cite{Pons:1998mm}, we decided
to implement the non-rotating profiles in an alternative
way.  As in the previous approach, we consider the spherical
configuration obtained by the evolution code at time $t$,
with central density $p_c$ and baryon mass (constant during
the evolution) $M_b$. To describe the rotating star, we use
the GM3 EoS $\epsilon=\epsilon(p,s,Y_L)$; since we are
restricting our analysis to slowly rotating
stars, the  entropy and lepton
fraction profiles $s(a)$ and $Y_L(a)$  of the non-rotating star are a good
approximation for those of the rotating star.
We  follow the steps discussed at the end of
Section~\ref{sec:HT}: (i) solve the TOV equations for a star with central
pressure $p_c+\delta p_c$; at each value of $a$, the energy density is
$\epsilon(p,s(a),Y_L(a))$; (ii) solve Hartle's perturbation equations, finding the
baryon mass of the star rotating to a given rate with this reduced
central pressure and find
the correction to the baryon mass due to rotation; (iii) iterate the
first two steps, finding $\delta p_c$ such that the baryon mass of the rotating
star is $M_b$.  We remark that the energy density of the rotating star in step
(ii) is related to that of the non-rotating star in step (i) by the
Hartle-Thorne prescription described above Eq.~(\ref{epspert}).

Since we are using an appropriate non-barotropic EoS, the instability discussed
above disappears, and the central pressure of the rotating star is, as expected,
smaller than that of the non-rotating star with same baryon mass.

We stress again that we are using the numerical solution of the transport
equations~\eqref{transpeq} for a {\it non-rotating} PNS, to build
quasi-stationary configurations of a {\it rotating} PNS. Therefore, we are
neglecting the effect of rotation on the time evolution of the PNS.  To be
consistent, we should have integrated the transport equations appropriate for a
rotating star, which are much more complicated. 
Since these approximations affect the timescale of the stellar evolution,
we would like to estimate how faster, or slower,
the rotating star looses its thermal and lepton content with respect to the 
non-rotating one.
\\
Since the evolution timescale is governed by neutrino diffusion
processes, at each time step of the non-rotating PNS evolution, we have computed
and compared the neutrino diffusion coefficients $D_n$ (see
Eqns.~\eqref{eq:Fnu}, \eqref{eq:Hnu}) for non-rotating and rotating
configurations.  The latter have been obtained by replacing the profiles
($p(a)$, $\epsilon(a)$, etc.) of a non-rotating PNS with those of a rotating PNS
(computed as discussed above in this Section).
In the upper and middle panels of Fig.~\ref{fig:compare} we plot $D_2, D_3$ and
$D_4$ as functions of the enclosed baryon mass $m_B=m_n a$, for the non-rotating
(solid line) and rotating (dashed line) configurations, at $t=0.2$ s, $t=1.2$ s
and $t=10$ s.
In the lower panels we plot the neutrino number density and the total
energy density at the same times.  We assume $M_b=1.6\,M_{\odot}$ and that
the initial angular momentum, $J_{in}$, is equal to the maximum angular momentum $J_{max}$, above which
mass-shedding sets in (see Sec.~\ref{sec:resspin} for further details). We see that  the
diffusion coefficients of the rotating configurations are  larger than those
of the non-rotating star.  
For $m_B\lesssim 1~M_\odot$ the relative difference $\vert
D_n^{rot}-D_n^{non~rot}\vert/\vert D_n^{non~rot}\vert$ is always smaller than
$\sim10-20\%$, and becomes  smaller than a few percent  after the first few
seconds.

In the outer region $m_B\gtrsim 1~M_\odot$ and early times, the relative
difference seems larger, in particular for the coefficient $D_3$, but this has
no effect for two reasons: first, as shown in the two lower panels of
Fig.~\ref{fig:compare}, both the neutrino number density and  the total energy
density are much  smaller than in the inner core; therefore, even though the
diffusion coefficients of the rotating star are larger than those of the
non-rotating one, few neutrinos are trapped in this region and transport effects
do not contribute significantly to the overall evolution; second, the
differences become large in the semi-transparent region, when the mean free path
becomes comparable to (or larger than) the distance to the star surface. In this
region the diffusion approximation breaks down and in practice the diffusion coefficients
are always numerically limited (a flux-limiter approach).

From the above discussion we can conclude that the
rotating star looses energy and lepton number through neutrino emission
faster than the non-rotating one. This effect is larger at the beginning of the
evolution, i.e. for $t \lesssim 2$ s, and is of the order of $\sim10-20\%$, 
but becomes negligible at later times. Consequently our rotating star cools
down and contracts over a timescale which, initially, is $\sim10-20\%$ shorter
than that of the corresponding non-rotating configuration.

\subsection{Evolution of the angular momentum and of the
rotation rate}\label{sec:momangev}
Once the equations describing the rotating  configuration
are solved for each value of the evolution time $t$ and for an assigned
value of the rotation  rate
$\Omega$,  the  solution of these equations allows one to  compute the
multipole moments of the rotating star,  including the angular momentum $J$. Conversely
we can choose, at  each value of $t$, the value of the angular momentum, and
determine, using a shooting method, the corresponding value of the rotation
rate.

If we want to describe the early evolution of a rotating PNS, we need a
physical prescription for  the time dependence of $J$.
For instance, we may  assume that the angular
momentum is constant, as
in~\cite{Villain:2003ey} (see also~\cite{Goussard:1996dp,Goussard:1997bn}).
However, in the first minute of a PNS life, neutrino
emission carries away
$\sim10\%$ of the star gravitational mass~\cite{Lattimer:2000nx}, and also a
significant fraction of the total angular momentum~\cite{Janka:2004np}.
To our knowledge, the most sensible estimate of the
neutrino angular momentum loss in PNSs has been done by Epstein
in~\cite{Epstein:1978md}
\begin{equation}
  \frac{dJ}{dt}=-\frac{2}{5}qL_\nu
R^2\Omega \label{epst}
\end{equation}
where $R$ is the radius of the star, $L_\nu=-dM/dt$ is
the neutrino energy flux, and $q$ is an efficiency parameter, which depends on
the features of the neutrino transport and emission. If neutrinos escape without
scattering, $q=1$; if, instead, they have a very short mean free path, they are
diffused up to the surface, and then are emitted with $q=5/3$. As discussed
in~\cite{Epstein:1978md} (see also \cite{Kazanas:1977nt,Mikaelian:1976ce,Henriksen:1978apj}),
$q=5/3$ should be considered as an {\it upper limit} of the angular momentum
loss by neutrino emission. A more recent, alternative
study~\cite{Dvornikov:2009rk} indicates  an angular momentum
emission smaller than this limit. In the following, we shall consider Epstein's formula
with $q=5/3$, and this  has to be meant as an upper limit. We also mention that a
simplified expression based on Epstein's formula for the angular momentum loss
in PNSs has been derived in~\cite{Janka:2004np} and used in~\cite{Martinon:2014uua}.

\subsection{Mass-shedding frequency}\label{sec:ms}
As mentioned in Sec.~\ref{sec:HT}, the perturbative approach which we use to model a rotating star is accurate up to
$\nu\lesssim0.8\nu_{ms}$, where $\nu_{ms}$ is the mass-shedding frequency.  The only quantity which is poorly estimated
is, of course, the mass-shedding frequency itself.  Therefore, $\nu_{ms}$ will be determined using a numerical fit
derived in~\cite{Doneva:2013zqa} from fully relativistic, non-linear integrations of Einstein's equations:
\begin{equation}
\nu_{ms}(Hz)=a\sqrt{\frac{M/M_\odot}{R/1{\rm km}}}+b\label{fitms}
\end{equation}
where $a=45862$ Hz and $b=-189$ Hz. We remark that the coefficients $a,b$ of this fit do not
depend on the EoS.

\subsection{Gravitational wave emission}\label{sec:gw}
If the evolving PNS is born with some degree of  asymmetry, it emits
gravitational waves. Assuming that the star rotates about a principal axis of
the moment of inertia tensor, i.e., that there is no precession\footnote{Free
precession requires the existence of a rigid crust~\cite{Jones:2000ud}, thus it
should not occur in the first tens of seconds of the PNS life, when the crust has not
formed yet~\cite{Suwa:2013mva}.}, gravitational waves are emitted at twice the orbital frequency
$\nu$, with amplitude
\cite{Zimmermann:1979ip,1987thyg.book..330T,Bonazzola:1995rb,Jones:2001ui}
\begin{equation}
h_0\simeq\frac{4G(2\pi\nu)^2I_3\epsilon}{c^4 r}\,.\label{gwamplitude}
\end{equation}
The deviation from axisymmetry is described by the
{\it ellipticity} $\epsilon$, defined as
\begin{equation}
\epsilon=\frac{I_1-I_2}{I_3}
\end{equation}
where $I_1$, $I_2$ and $I_3$ are the principal moments of inertia of the
PNS and $I_3$ is assumed to be aligned with the rotation axis.
For old neutron stars, the loss of energy through gravitational waves
is compensated by a decrease of rotational energy,
which contributes to the spin-down of the star (the main contribution to the spin-down 
being  that of the magnetic field).

In the case of a newly born PNS the situation is different.
As the star contracts, due to the processes related to neutrino production and
diffusion, its rotation rate increases. If the PNS has a finite ellipticity,
it emits gravitational waves, whose amplitude and frequency also
increase as the star spins up. The timescale of this process is of the order of
tens of seconds. In our model, for simplicity we shall assume that the PNS
ellipticity remains constant over this short time interval.


Unfortunately, the ellipticity of a PNS is  unknown. In cold,
old NSs it is expected to be, at most, as large as
$\sim10^{-5}-10^{-4}$~\cite{Haskell:2006sv,Ciolfi:2010td} (larger values are
allowed for EoS including exotic matter
phases~\cite{Horowitz:2009ya,JohnsonMcDaniel:2012wg}).
For newly born PNSs, it may be larger, but we have no hint on its actual value.
To our knowledge, current numerical simulations of core-collapse do not provide
estimates of the PNS ellipticity. We remark that although there is observational
evidence of large asymmetries in supernova
explosions~\cite{Wang:2002nc,Leonard:2006qy}, there is no evidence that
they can be inherited by the PNS. 
 In the following, we shall assume
$\epsilon=10^{-4}$, but this should be considered as a fiducial value: the
gravitational wave amplitude (which is linear in $\epsilon$) can be easily
rescaled for different values of the PNS ellipticity.

\section{Results}\label{sec:results}
\subsection{Spin evolution of the proto-neutron star}\label{sec:resspin}
\begin{figure}[ht]
\begin{center}
\leftskip-4em
\epsfig{file=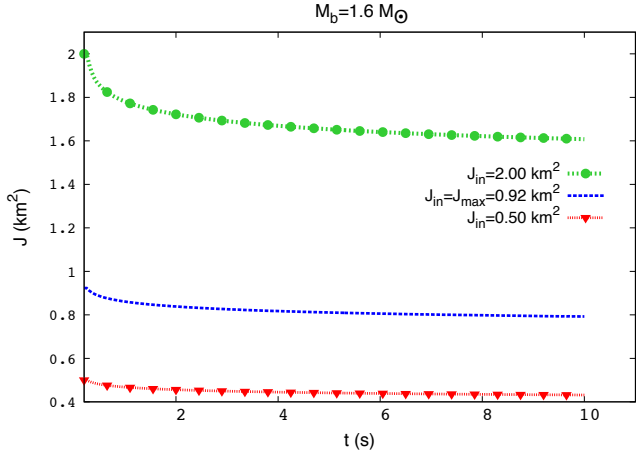,width=9.5cm,clip=true}
\caption{Angular momentum evolution due to neutrino losses, for a PNS with baryonic mass $M_b=1.6\,M_\odot$ and initial
  angular momentum $J_{in}=(2.02,\,3.71,\,8.08)\times10^{48}$erg s.  \label{fig:Jt}}
\end{center} \end{figure}

In Figure~\ref{fig:Jt} we show how the angular momentum changes according to
Epstein's formula~\eqref{epst} as the PNS evolves. We assume $q=5/3$ and
baryonic mass $M_b=1.6\,M_\odot$. We consider different values of the angular
momentum $J_{in}$ at the beginning of the quasi-stationary phase ($t=0.2$ s
after the bounce): $J_{in}=\unit[2.02\times10^{48}]{erg\,s}$,
$J_{in}=\unit[3.71\times10^{48}]{erg\,s}$ and
$J_{in}=\unit[8.08\times10^{48}]{erg\,s}$. We find that, in the
first ten seconds after bounce, $13\%$ of the initial angular momentum is
carried away by neutrinos if $J_{in}=\unit[2.02\times10^{48}]{erg\,s}$ or
$J_{in}=\unit[3.71\times10^{48}]{erg\,s}$; $20\%$ of the initial angular
momentum is carried away if $J_{in}=\unit[8.075\times10^{48}]{erg\,s}$.
As mentioned above, $q=5/3$ should be considered as an upper bound; for smaller
values of $q$, the rate of angular momentum loss would be smaller.

\begin{figure}[ht]
\begin{center}
\leftskip-4em
\epsfig{file=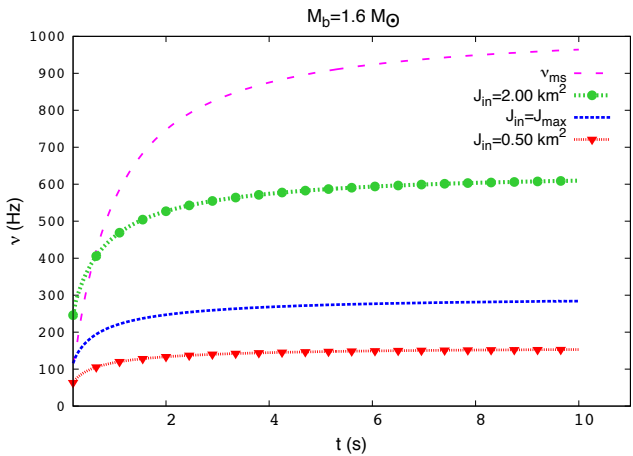,width=9.5cm,clip=true}
\caption{Evolution of the PNS rotation rate, corresponding to the angular momentum profiles
shown in Figure~\ref{fig:Jt}.  \label{fig:nut}} \end{center} \end{figure}

The corresponding evolution of the PNS rotation frequency is shown in Figure~\ref{fig:nut}. In the same Figure we also
show the mass-shedding frequency $\nu_{ms}$, computed using the fit~\eqref{fitms}. We see that if
$J_{in}=\unit[8.08\times10^{48}]{erg\,s}$, the curves of $\nu(t)$ and of $\nu_{ms}(t)$ cross during the quasi-stationary
evolution; before the crossing, the PNS spin is larger than the mass-shedding limit.  This means that a PNS with such
initial angular momentum would lose mass.  If we require the initial rotation rate to be smaller than the mass-shedding
limit, we must impose $J_{in} \le J_{max}\equiv\unit[3.72\times10^{48}]{erg\,s}$.  We remark that the value of $J_{max}$
is not affected by the efficiency of angular momentum loss $q$: if $q<5/3$, $J_{max}$ has the same value, but the
rotation rate grows more rapidly than in Figure~\ref{fig:nut}.

It is interesting to note that, since $\nu_{ms}$ has a steeper increase than $\nu(t)$, even when the bound
$\nu\le\nu_{ms}$ is saturated at the beginning of the quasi-stationary phase the frequency becomes much smaller than the
mass shedding frequency at later times. This is an a-posteriori confirmation that the slow rotation approximation is
appropriate to study newly born PNSs.  For $t>10$ s, the PNS radius does not change significantly, and the star starts
to spin down due to electromagnetic and gravitational emission. However this spindown timescale is much longer than the
timescale of the quasi-stationary evolution we are considering; therefore it is unlikely that after this early phase the
PNS rotation rate is larger than $\sim 300$ Hz (i.e., that its period is smaller than
$\sim3$ ms), unless some spin-up mechanism (such as e.g. accretion) sets in. A less efficient angular momentum loss
($q<5/3$) would moderately increase this final value, but the general picture would remain the same.

It is worth noting that models of
pre-supernova stellar evolution~\cite{Heger:2004qp} predict a similar range of the PNS rotation rate and angular
momentum. Among the models considered in~\cite{Heger:2004qp}, the only one with $J>J_{max}$ (and rotation period smaller
than $3$ ms) is expected to collapse to a black hole.  Other works~\cite{Thompson:2004if,Ott:2005wh} have shown that
if the progenitor has a rotation rate sufficiently large, the PNS resulting from the core-collapse can have periods of
few ms; our results suggest that this scenario is unlikely, unless there is a significant mass loss in the early
Kelvin-Helmoltz phase.

\subsection{Gravitational wave emission}
\begin{figure}[ht]
\begin{center}
\leftskip-2em
\epsfig{file=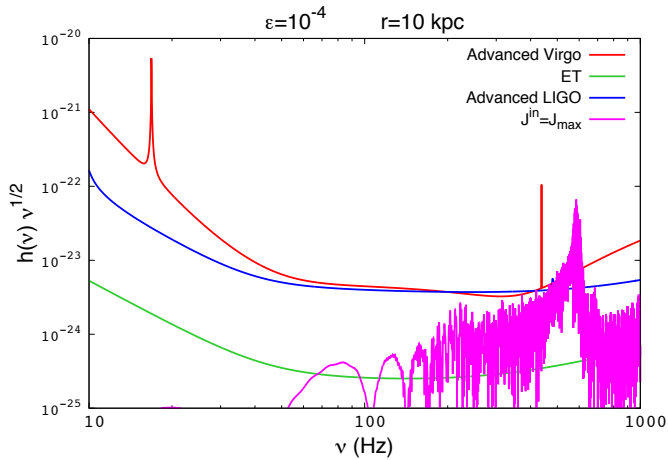,width=9.5cm,clip=true}
\caption{
The strain amplitude $\tilde h(f)\sqrt{f}$
of the gravitational wave signal emitted by a PNS with 
$\epsilon=10^{-4}$, $J_{in}=J_{max}$, at a distance $r=10$ kpc,
is compared with the noise curves of Advanced Virgo, Advanced LIGO and ET.
\label{fig:h}}
\end{center}
\end{figure}

If the PNS has a finite ellipticity $\epsilon$ (which we assume, for simplicity,
to remain constant during the first $\sim10$ s of the PNS life), it emits
gravitational waves with frequency $f(t)=2\nu(t)$ and amplitude given by
Eq.~\eqref{gwamplitude},

\begin{equation}
h_0\simeq\frac{4G(2\pi\nu(t))^2I_3(t)\epsilon}{c^4r}\,.\label{gwamplitude2}
\end{equation}
As the spin rate $\nu(t)$ increases, both the frequency and the amplitude of the
gravitational wave increase; therefore, the signal is a sort of ``chirp''; this
is different from the chirp emitted by neutron star binaries before coalescence,
because the amplitude increases at a much milder rate.
In Figure~\ref{fig:h} we show the strain amplitude $\tilde
h(f)\sqrt{f}=\sqrt{f}\sqrt{(\tilde h_+(f)^2+\tilde h_\times(f)^2)/2}$, 
where $\tilde h_{+,\times}(f)$ are the Fourier transform of the two polarization
of the gravitational wave signal
\begin{eqnarray}
h_+&=& h_0\frac{1+\cos^2i }{2}\cos(2\pi f(t) t)\\
h_\times&=& h_0\cos i\sin(2\pi f(t) t),
\end{eqnarray}
and $i$ is the angle between the rotation axis and the line of sight.
In Figure~\ref{fig:h} the signal strain amplitude,  computed assuming optimal
orientation, $J_{in}=J_{max}$, $\epsilon=10^{-4}$ and a distance of
$r=10$ kpc,  is
compared with the sensitivity curves of Advanced Virgo\footnote{https://inspirehep.net/record/889763/plots},
Advanced LIGO\footnote{https://dcc.ligo.org/LIGO-T0900288/public}, and of the
third generation detector ET\footnote{http://www.et-gw.eu/etsensitivities}.
We see that the signal is marginally above noise
for the advanced detectors, but it is definitely above the noise curve for ET.
This signal would be seen by Advanced Virgo with a signal-to-noise ratio
$SNR=1.4$, and by Advanced LIGO with $SNR=2.2$, too low to extract it
from the detector noise; however, since the
signal-to-noise ratio scales linearly with the ellipticity, a star born with
$\epsilon=10^{-3}$ would be detected with $SNR=14$ and $SNR=22$ by Advanced
Virgo and LIGO, respectively.
The third generation detectors like ET would detect the signal coming from a
galactic PNS born with $\epsilon=10^{-4}$ with a very large signal-to-noise
ratio, i.e.  $SNR=22$. If the source is in the Virgo cluster ($d=15~Mpc$), the
ellipticity of the PNS should be as large as $5\cdot 10^{-2}$ to be seen by ET with
$SNR=8$.

\section{Concluding remarks} \label{sec:conclusions}
In this paper we have studied the angular momentum loss, the time dependence of the rotation rate and the gravitational
wave emission of a newly born PNS, during the first tens of seconds after bounce.  The early evolution of the rotating
PNS has been modeled using the entropy and lepton fraction profiles  consistently
computed solving the general relativistic transport equations for a non-rotating star; angular momentum loss due to
neutrino emission has been modeled using Epstein's formula~\cite{Epstein:1978md}.

During this early evolution, the star spins up due to contraction.
By requiring that the
initial rotation rate does not exceed the mass-shedding limit, we
have estimated the maximum rotation rate at the end of the spin-up phase. For a
PNS of $M_b=1.6\,M_\odot$ we  find  that  one minute
after bounce the star would rotate at  $\nu\lesssim 300$ Hz, corresponding to
a rotation period $\tau_{min}\gtrsim\unit[3.3\times10^{-3}]{s}$.

If the PNS is born with a finite ellipticity $\epsilon$, while spinning up it emits gravitational
waves at twice the rotation frequency. This signal increases
both in frequency and amplitude. We find that
for a galactic supernova, if $\epsilon=10^{-3}$ this signal could be detected by Advanced
LIGO/Virgo with a signal-to-noise ratio $\gtrsim 14$.
To detect farther sources, third generation detectors like ET would be needed.

We remark that  the actual value of PNS ellipticities is unknown, and depends on
the details of the supenova core collapse.
Accurate numerical simulations of supernova explosion addressing this
issue are certainly needed to  provide a quantitative estimate of
the range of $\epsilon$.

We also remark that in our approach the effects of the PNS rotation are
consistently included in the structure equations, but they are neglected when
solving the neutrino transport equations. We estimate that due to this
approximation, we overestimate the evolution timescale at early times of, at
most, $\sim10-20\%$.  Moreover, since we are not interested in the details of
the neutrino dynamics and we need a fast code to evolve the star for tens of
seconds, we perform energy averages to determine the neutrino diffusion
coefficients, and we apply a flux-limiter; these approximations should not
significantly affect the thermodynamical evolution of the PNS and its
gravitational wave emission.

This work is a first step in the study of the early evolution of PNSs.
A paper with a detailed description of our numerical code, and its extension to more
recent EoSs, is in preparation \cite{Camelio:2016}. Further
developments shall include differential rotation,  convection
and generalization of the neutrino transport equations to rotating
PNSs.

\begin{acknowledgments}
We thank O. Benhar and A. Lovato for useful discussions on the EoS of
the PNS. We also thank S. Reddy and L.F. Roberts for useful discussions on the
neutrino cross sections.
This work was partially supported by ``NewCompStar'' (COST Action MP1304),
and by the H2020-MSCA-RISE-2015 Grant No. StronGrHEP-690904.
J.A.P. acknowledges support by the MINECO grants AYA2013-42184-P and AYA2015-66899-C2-2-P.
\end{acknowledgments}

\appendix
\section{Hartle-Thorne equations}\label{app:HT}
Here, we briefly describe the equations of the perturbative Hartle-Thorne approach discussed in Sec.~\ref{sec:HT}. For
further details we refer the reader to~\cite{Hartle:1967he,Hartle:1968si,hartle1973slowly} and to the Appendix
of~\cite{Benhar:2005gi}.

The spacetime metric (up to order $O(\Omega^3)$) is given by Eq.~(\ref{metricHartle}); it depends on the background
functions $\phi(r)$, $\lambda(r)=-\log(1-2m(r)/r)$, and on the perturbations functions $h_l(r)$, $m_l(r)$ ($l=0,2$), $k_2(r)$,
$\omega(r)$, $w_l(r)$ ($l=1,3$). The energy and pressure (Eulerian) perturbations are
\begin{eqnarray}
\delta P&=&(\epsilon(r)+P(r))(\delta p_0(r)+\delta p_2(r)P_2(\mu))\nonumber\\
\delta\epsilon&=&\frac{d\epsilon/dr}{dP/dr}\delta P\,,
\end{eqnarray}
and depend on the perturbation functions $p_l(r)$ ($l=0,2$).

The background spacetime is described by the TOV equations:
\begin{eqnarray}
  \frac{dm}{dr}&=&4\pi r^2\epsilon\nonumber\\
  \frac{d\phi}{dr}&=&2\frac{m+4\pi r^3P}{r(r-2m)}\nonumber\\
  \frac{dP}{dr}&=&-\frac{\epsilon+P}{2}\frac{d\phi}{dr}\,.
\end{eqnarray}
The mass of the non-rotating configuration is obtained by matching at the stellar surface $r=R$ the interior solution
with the exterior (Schwarzschild) solution, i.e., $M=m(R)$. Moreover, the baryonic mass $M_b$ of the non-rotating
configuration is obtained integrating the equation $dm_b/dr=4\pi r^2 e^{\lambda/2}\rho$, and computing $M_b=m_b(r)$.

The spacetime perturbation to first order in $\Omega$ is described by the function $\omega(r)$, which is responsible for the
dragging of inertial frames; it satisfies the equations
\begin{eqnarray}
  \frac{d\chi}{dr}&=&\frac{u}{r^4}-\frac{4\pi r^2(\epsilon+P)\chi}{r-2M}\\
  \frac{du}{dr}&=&\frac{16\pi r^5(\epsilon+P)\chi}{r-2M}\,,
\label{eqpai}
\end{eqnarray}
where $\varpi=\Omega-\omega$, $j(r)= e^{-\phi/2}(1-2M/r)^{1/2}$, $\chi=j\varpi$ and $u=r^4jd\varpi/dr$. The angular
momentum $J$ is obtained by matching the interior with the exterior solution $\chi(r)=\Omega-2J/r^3$, $u(r)=6J$ at
$r=R$. The moment of inertia, at zero-th order in the rotation rate, is $I=J/\Omega$.

The perturbations to second order in $\Omega$ are described by the metric functions $h_l(r)$, $m_l(r)$ ($l=0,2$),
$k_2(r)$, and by the fluid pressure perturbations $\delta p_l$. The $l=0$ perturbations satisfy the equations
\begin{eqnarray}
\frac{d}{dr}\left(\delta p_0+h_0-\frac{\chi^2r^3}{3(r-2M)}\right)&=&0\nonumber\\
\delta p_2+h_2-\frac{\chi^2r^3}{3(r-2M)}&=&0\nonumber\\
\end{eqnarray}
and
\begin{eqnarray}
\frac{dm_0}{dr}&=&4\pi r^2\frac{d\epsilon}{dP}[\delta p_0(\epsilon+P)]+
\frac{u^2}{12r^4}+\frac{8\pi r^5(\epsilon+P)\chi^2}{3(r-2M)}\nonumber\\
\frac{d\delta p_{0}}{dr}&=&\frac{u^2}{12r^4(r-2M)}-\frac{m_0(1+8\pi r^2P)}{(r-2M)^2} \nonumber\\
&&-\frac{4\pi(\epsilon+P)r^2\delta p_0}{r-2M}+\frac{2r^2\chi}{3(r-2M)}\left[\frac{u}{r^3}\right.\nonumber\\
&&\left.+\frac{(r-3M-4\pi r^3P)\chi}{r-2M}\right]\,.
\end{eqnarray}
Matching the interior and the exterior solutions at $r=R$, it is possible to compute the correction to the mass due to
stellar rotation, $\delta M=m_0(R)+J^2/R^3$, and the monopolar stellar deformation. The baryonic mass correction $\delta
M_b=\delta m_b(R)$ is given by solving the equation
\begin{eqnarray}
\frac{d\delta m_b}{dr}&=&4\pi r^2e^{\lambda/2}\left[\left(1+\frac{m_0}{r-2m}+\frac{1}{3}
r^2\varpi^2e^{-\phi}\right)\epsilon\right.\nonumber\\
&&\left.+\frac{d\epsilon/dr}{dP/dr}(\epsilon+P)\delta p_0\right]\,.
\end{eqnarray}
The $l=2$ perturbations satisfy the equations
\begin{eqnarray}
\frac{dv_2}{dr}&=&-\frac{d\phi}{dr}h_2+(\frac{1}{r}+
\frac{1}{2}\frac{d\phi}{dr})\left[
\frac{8\pi r^5(\epsilon+P)\chi^2}{3(r-2M)}+\frac{u^2}{6r^4}\right]\nonumber\\
\frac{dh_2}{dr}&=&\left[-\frac{d\phi}{dr}+
\frac{r}{r-2M}(\frac{d\phi}{dr})^{-1}(8\pi(\epsilon+P)-
\frac{4M}{r^3})\right]h_2\nonumber\\
&-&\frac{4v_2}{r(r-2M)}(\frac{d\phi}{dr})^{-1}+
\frac{u^2}{6r^5}\left[
\frac{1}{2}\frac{d\phi}{dr}r-\frac{1}{r-2M}(\frac{d\phi}{dr})^{-1}\right]\nonumber\\
&+&\frac{8\pi r^5(\epsilon+P)\chi^2}{3(r-2M)}\left[
\frac{1}{2}\frac{d\phi}{dr}r+
\frac{1}{r-2M}(\frac{d\phi}{dr})^{-1}\right]\,,
\end{eqnarray}
where $v_2=k_2+h_2$.
Matching the interior and exterior solutions, it is possible to determine the quadrupole moment of the PNS and its
quadrupolar deformation.

The equations for the peturbations at $O(\Omega^3)$, $w_l(r)$ ($l=1,3$), have a similar structure but they are longer
and are not reported here; we refer the reader to~\cite{hartle1973slowly,Benhar:2005gi}. They yield the octupole moment,
the third-order corrections to the angular momentum and the second-order corrections to the moment of inertia.

\bibliography{biblio}
\end{document}